\documentclass{aa}
\usepackage[varg]{txfonts}
\usepackage[utf8]{inputenc}
\usepackage{comment}
\usepackage{amsmath}
\usepackage{graphicx}
\usepackage{gensymb}
\usepackage[colorlinks=true,linkcolor=blue,citecolor=blue]{hyperref}
\usepackage{xspace}
\usepackage{geometry}  
\usepackage{lipsum}    

\usepackage{gensymb}
\DeclareRobustCommand{\ion}[2]{\textup{#1\,\textsc{\lowercase{#2}}}}
\newcommand{\jwst}{\emph{JWST}\xspace}

\newcommand{\sirocco}{\textsc{Sirocco}\xspace}

\newcommand{\hei}{\ion{He}{i}\xspace}
\newcommand{\heii}{\ion{He}{ii}\xspace}

\usepackage{orcidlink}
\usepackage{natbib}
\bibpunct{(}{)}{;}{a}{}{,} 

\begin{document}

\title{The nature of `little blue dots'}

\author{Albert Sneppen\inst{\ref{addr:DAWN},\ref{addr:jagtvej},\ref{addr:IoA}}\orcidlink{0000-0002-5460-6126},
Darach Watson\inst{\ref{addr:DAWN},\ref{addr:jagtvej}}\orcidlink{0000-0002-4465-8264},
James H. Matthews\inst{\ref{addr:oxford}}\orcidlink{0000-0002-3493-7737} and 
Georgios Nikopoulos\inst{\ref{addr:DAWN},\ref{addr:jagtvej}}\orcidlink{0009-0004-6791-9246}
}

\authorrunning{Sneppen et al.}

\institute{Cosmic Dawn Center (DAWN)\label{addr:DAWN}
\and
Niels Bohr Institute, University of Copenhagen, Jagtvej 128, DK-2200, Copenhagen N, Denmark \label{addr:jagtvej} 
\and
Institute of Astronomy, University of Cambridge, Madingley Road, Cambridge, CB3 0HA, UK \label{addr:IoA}
\and Astrophysics, Department of Physics, University of Oxford, Keble Road, 
Oxford OX1 3RH, UK\label{addr:oxford}
}


\abstract{
Previous \sirocco radiative-transfer models of gas-cocooned AGN predicted lower-column counterparts to little red dots (LRDs): compact, X-ray-weak sources with bluer continuum slopes and Balmer jumps rather than Balmer breaks. The recently identified population of little blue dots (LBDs) closely resembles this predicted phase.
Here we explore these lower-column-density cocoons in which nebular recombination emission remains visible while strong Balmer-continuum absorption is avoided. We find that a sequence of increasing column density connects more classical AGN spectra, Balmer-jump LBD-like spectra at $N_{\rm H}\!\sim\!{\rm few}\times10^{24}\,\mathrm{cm^{-2}}$, and Balmer-break LRD-like spectra at higher columns. Along this sequence, electron scattering produces exponential line wings and suppresses X-ray emission before strong Balmer absorption features emerge. 
The models predict that LBDs should share X-ray weakness and exponential line wings with LRDs, while being distinguished by Balmer jumps in emission, weak or absent absorption features, potential \ion{He}{ii}\,$\lambda$4686 emission, lower H$\alpha$ luminosities and narrower H$\alpha$ lines. We compare with spectra of eight LBDs and find that significant Balmer-jump signatures are common. We therefore suggest that LBDs are lower-column analogues of LRDs within a common gas-cocooned AGN sequence.
}
\maketitle

\section{Introduction}
An early phase of supermassive black hole (SMBH) growth buried in translucent and dense gas has been invoked to explain the \jwst population of `little red dots' \citep[LRDs,][]{Matthee2024}. Notable signatures of the gas include continuum and line absorption from hydrogen gas in $n\!=\!2$ \citep{Setton2024,Inayoshi2025,deGraaff2025,Naidu2025,Torralba2026}, deviations from `Case~B' line ratios \citep{Nikopoulos2025,DEugenio2025b,Torralba2025} and exponential wings indicative of an electron-scattering cocoon \citep{Rusakov2025,Chang2025,Kokorev2025,Sneppen2026,Matthee2026} that could be responsible for partially suppressing X-ray emission due to the Compton-thick gas \citep{Maiolino_Chandra_2024,Comastri2026,Tortosa2026,Sneppen2026c}. Self-consistent radiative transfer modelling of LRDs  \citep{Sneppen2026} predicts a complementary population of lower-column cocooned AGN, which should be characterised by exponential line profiles and Balmer continuum emission (i.e.\ a Balmer jump as opposed to a break).


A high-redshift AGN population recently dubbed `little blue dots' \citep[LBDs,][]{Brazzini2026} is distinguished by their blue rest-frame UV--optical continua, but, similar to LRDs, LBDs display compact morphologies, non-detection of X-rays, weak hot dust emission and exponential-shaped broad lines \citep{Rusakov2025,Brazzini2026,Scholtz2026,Geris2026}. This phenomenological overlap suggests that LBDs may represent a more weakly dust-reddened view of the same accretion phase due to orientation \citep{Madau2026}, or they may be the lower-column cocooned AGN inferred by \cite{Matthee2026} and predicted by radiative transfer modelling of ionised gas cocoons \citep{Sneppen2026,Sneppen2026b}. A key distinction between the two hypothesised gas-cocoon regimes is the visibility of the Balmer jump emission. In the higher column density \sirocco models, with $N_{\rm H}\sim10^{25}\,\mathrm{cm^{-2}}$, strong nebular emission signatures are produced, including larger H$\alpha$ luminosities and a Paschen jump, but the associated Balmer jump is partly hidden by absorption in a partially neutral layer at larger radial distances \citep{Sneppen2026b}. This absorption in an outer layer produces the Balmer-continuum inflection and line absorption characteristic of LRDs. By contrast, the blue continua of LBDs suggest a more direct view of the Balmer-jump region, in which nebular recombination emission remains visible and without strong imprinted neutral absorption.

Here we investigate the lower-column, Balmer-jump phase of cocooned AGN and examine whether it can account for the emerging LBD population, focusing on their predicted continua, exponential broad-line wings, X-ray weakness, and connection to LRDs.

\begin{figure}
\vspace{-0.2cm}
\begin{center}
    \includegraphics[angle=0,width=0.49\textwidth]{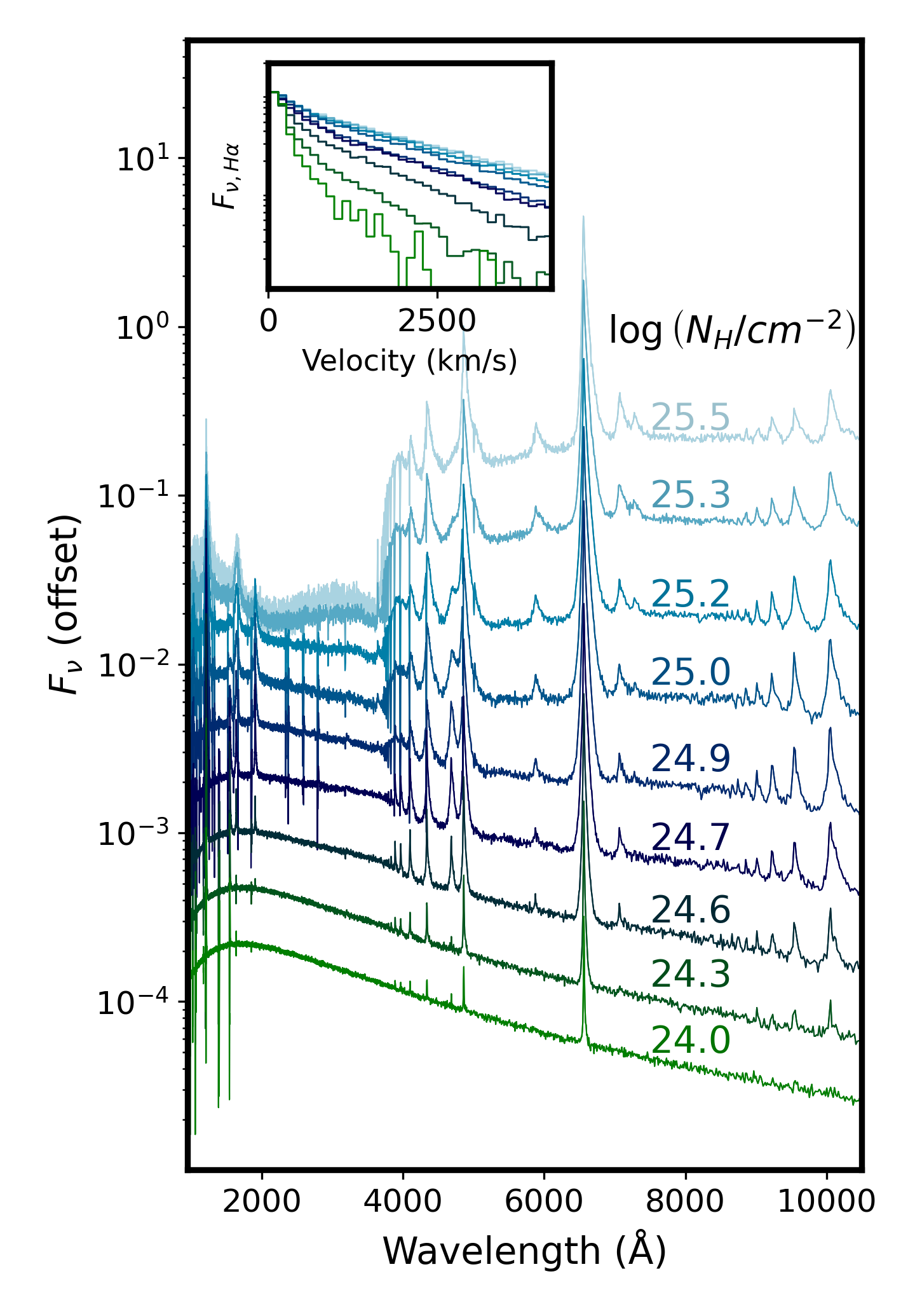}
\end{center}
\vspace{-0.9cm}
\caption{
\sirocco model sequence at fixed
\(R_{\rm in}=10^{16}\,\mathrm{cm}\), varying
\(n_{\rm H}(R_{\rm in})\) and hence the integrated column
\(N_{\rm H}\). Spectra are vertically offset for clarity. At low columns, $N_{\rm H}\lesssim10^{24}\,\mathrm{cm^{-2}}$, the emergent spectrum remains close to the assumed incident SED. At intermediate columns, $N_{\rm H}\sim10^{24.5}\,\mathrm{cm^{-2}}$, nebular Balmer-continuum emission produces LBD-like Balmer jumps. At higher columns, $N_{\rm H}\sim10^{25}\,\mathrm{cm^{-2}}$, neutral absorption produces LRD-like Balmer breaks. The inset shows the corresponding H$\alpha$ profiles, demonstrating that exponential electron-scattering wings arise across both the LBD-like and LRD-like column regimes, and grow wider with increasing column density in this regime.}
\label{fig:dens_seq}
\end{figure}

\begin{figure*}
\vspace{-0.1cm}
\begin{center}
    \includegraphics[angle=0,width=\textwidth]{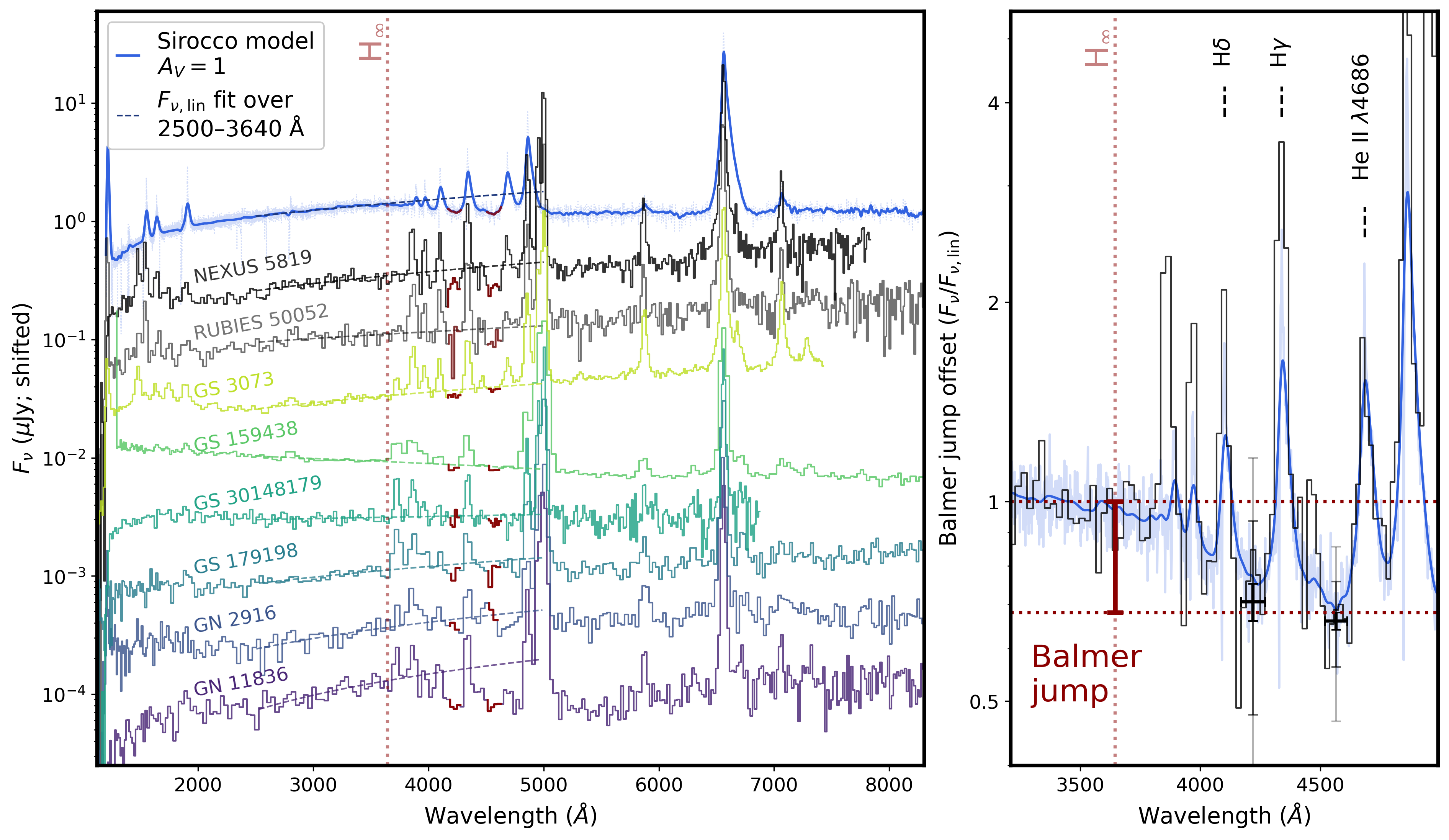}
\end{center}
\vspace{-0.5cm}
\caption{Comparison of a set of LBD spectra to a lower column density \sirocco model spectrum (light-blue at \sirocco model's resolution and dark-blue at NIRSpec/PRISM resolution). Left: UV and optical spectra with strong permitted H and He line emission. Right: zoom-in on the Balmer-limit region, where the same \sirocco model and Nexus\,5819 continua show a Balmer-jump-like deficit redward of the Balmer limit. $F_{\nu,\rm lin}$ is the linear fit to the continuum blueward of the Balmer edge ($\lambda\in[2500\,\AA,3640\,\AA]$), which highlights the continuum offset across H$_\infty\!=\!3645$\,\AA. Within the wavelength-windows least affected by line emission in the model space (highlighted with red), we show the mean observed flux-ratio with $1$, $5$ and $10\sigma$ statistical uncertainty indicated, showing that the observed flux ratios are significantly below unity, as expected for a Balmer-jump offset.   }
\label{fig:Balmer_jump_com}
\end{figure*}

\section{Methods}\label{sec:model}
We model cocooned AGN using the Sobolev Monte Carlo code \sirocco \citep{long2002,Matthews2025}, which follows photon propagation through multi-dimensional, optically thick flows and includes electron scattering, Sobolev line transfer, and bound-free processes. For a specified gas density and velocity structure, \sirocco self-consistently computes the radiation field and ionisation state, enabling predictions of continuum shapes, line profiles, line ratios and luminosities, and their correlations across the spectral energy distribution (SED) from X-rays through the near-infrared \citep{Sneppen2026b,Sneppen2026,Sneppen2026c}.

The models considered here are cocoons with hydrogen column densities of $N_\mathrm{H}=\int n_{\rm H}(r)\,{\rm d}r\!\sim\!\text{few}\times10^{24}\,\mathrm{cm^{-2}}$ with a local hydrogen number density following $n_{\rm H}(r)=n_{\rm H}(R_{\rm in})\,\left({r}/{R_{\rm in}}\right)^{-2}$ with inner and outer cocoon radii, $R_{\rm in}\in10^{15.5}$-$10^{17}\,{\rm cm}$ and \(R_{\rm out}=10\,R_{\rm in}\), sub-solar metallicity, $Z=0.1\,Z_{\odot}$, characteristic velocities of $v\!\sim\!\text{few}\times100\,\mathrm{km\,s^{-1}}$, illuminated by a central blackbody source with $T_{\rm BB}=30\,000$\,K following the models in \cite{Sneppen2026}. 
Fig.~\ref{fig:dens_seq} shows one-dimensional models in a density-sequence (i.e., $N_\mathrm{H}=(1{-}15)\times10^{24} \,{\rm cm^{-2}}$), where an individual spectrum is further detailed in Fig.~\ref{fig:Balmer_jump_com} with $N_\mathrm{H}=5\times10^{24} \,{\rm cm^{-2}}$. In Figs.~\ref{fig:Balmer_jump_pan} and \ref{fig:HeII}, we employ a 2-dimensional model with a similar column density and inflows/outflows from \cite{Sneppen2026}. Here we also adopt a narrow-line Seyfert~1 input X-ray spectrum with $k_{\rm bol,X}=L_{\rm bol}/(L_{2-10\,\mathrm{keV}})=50$ to examine Compton-thick X-ray suppression using the same \sirocco-based treatment as in \cite{Sneppen2026c}. Even for the 2-dimensional model we keep a cocoon covering fraction of unity (e.g., as motivated by the high-covering fraction inferred by \cite{Yanagisawa2026}), although given the multi-dimensional setup this could naturally be varied within \sirocco in future work. For Figs.~\ref{fig:Balmer_jump_com}-\ref{fig:HeII}, we convolve the model spectra to the observed spectral resolution and modestly modify the continuum UV--optical slope with reddening of $A_V=1$ using a Calzetti attenuation prescription \citep{Calzetti2000} without imposing the very steep far-UV suppression characteristic of the SMC extinction curve. This reddening is somewhat larger than, but of the same order as, that inferred for GS\,3073 with $A_V\sim0.4{-}0.5$ \citep{Ji2024} and is consistent with the generally weak mid-infrared dust emission of the LRD class \citep{Chen2025}. 
Fig.~\ref{fig:dens_seq} fixes $R_{\rm in}$ and varies the local density $n_H$, thereby changing \(N_{\rm H}\). Conversely, Figs.~\ref{fig:NH}-\ref{fig:beta} illustrate variations of both local density and geometrical scale. We organise the sequences by \(N_{\rm H}\), because it most directly controls the integrated electron-scattering and continuum optical depths.

These model spectra are typical of this part of the simulation grid, rather than extreme cases. For these gas columns, the gas is moderately optically thick to electron scattering, while the partially neutral absorbing layer remains weak enough that strong Balmer-continuum and Balmer-line absorption are avoided.

\subsection*{Observational sample}
\begin{figure*}
\vspace{-0.1cm}
\begin{center}
    \includegraphics[angle=0,width=\textwidth]{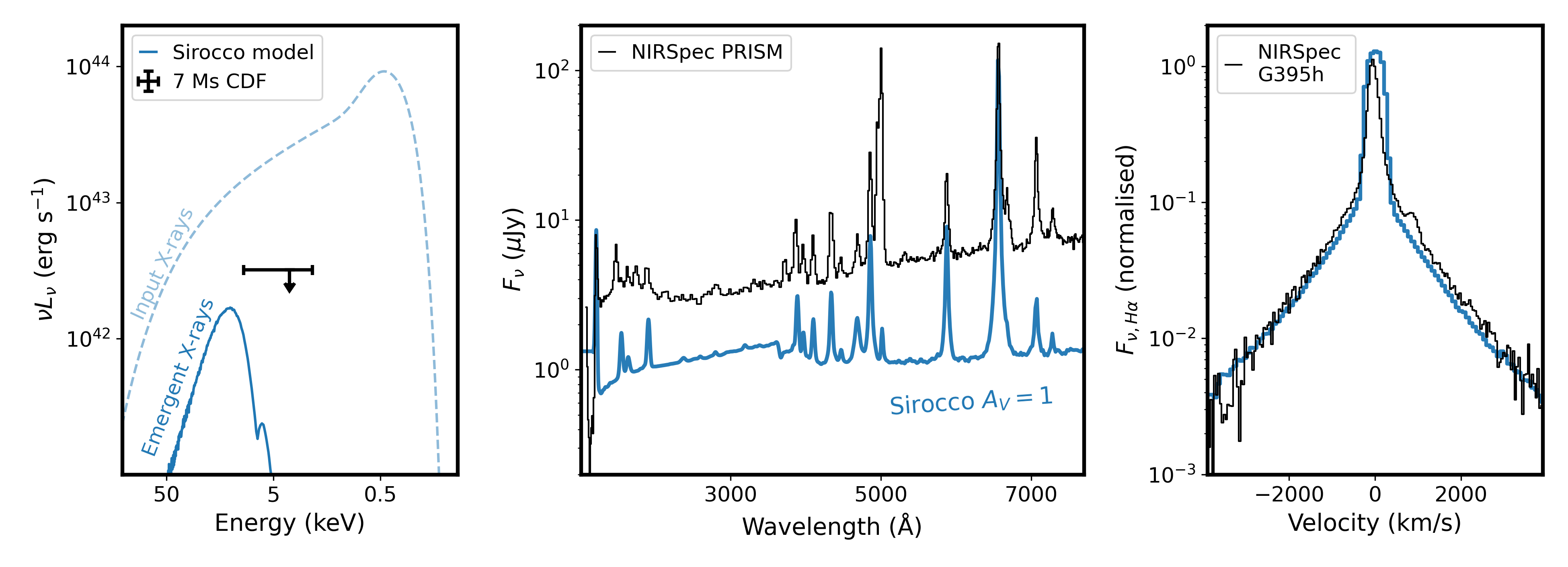}
\end{center}
\vspace{-0.8cm}
\caption{Panchromatic comparison between a single \sirocco model and GS\,3073 observational constraints \citep{Ubler2023}. \emph{Left}: Emergent X-ray spectrum compared with the intrinsic narrow-line Seyfert~1--like input template, assuming $k_{\rm bol,X}=50$, and with the 7\,Ms Chandra Deep Field constraint \citep{Maiolino_Chandra_2024}. The Compton-thick cocoon suppresses the emergent X-ray luminosity below the observational limit. \emph{Centre}: rest-frame UV–optical spectra, showing bound-free emission and strong recombination lines. \emph{Right}: H$\alpha$ profile with broad symmetric exponential wings due to electron-scattering gas. The same electron-scattering column that suppresses the X-ray emission also broadens H$\alpha$.   }
\label{fig:Balmer_jump_pan}
\end{figure*}

To illustrate the predicted Balmer-jump behaviour, we select LBDs with sufficiently high-S/N spectroscopy and continuous rest-frame coverage across the Balmer limit. This includes Nexus\,5819 at \(z=5.76\) \citep{Shen2024,Zhuang2026}, Rubies 50052 at \(z=5.24\) \citep[previously presented as CEERS\,02782 or CEERS 1670;][]{Kocevski2023_first,Harikane2023,Brooks2024} and GS 3073 at \(z=5.55\) \citep{Ubler2023,Ji2024,Brazzini2026,Matthee2026}. We additionally include the secure compact LBDs GS 159438 (\(z=3.24\)), GN 2916 (\(z=3.67\)), GS 179198 (\(z=3.83\)), GN 11836 (\(z=4.41\)) and GS 30148179 (\(z=5.92\)) from \cite{Geris2026}. These comprise all non-tentative \cite{Geris2026} LBDs with continuous wavelength coverage and a median PRISM S/N \(>3\). The selection does not depend on the measured Balmer-jump ratio and is intended for comparison with the model predictions rather than as a complete observational census. Nexus\,5819 is used for the detailed Balmer-jump comparison in Fig.~\ref{fig:Balmer_jump_com} because its offset is similar to that of the illustrative model, whereas GS 3073 is used for the panchromatic comparison (Figs.~\ref{fig:Balmer_jump_pan} \& \ref{fig:HeII}) because it has particularly deep X-ray constraints and high-resolution Balmer-line spectroscopy.

\section{Results}\label{sec:results}

At column densities lower than those required to model LRDs (see Fig.~\ref{fig:dens_seq}), \sirocco predicts a population of cocooned AGN with nebular emission signatures indicated by a Balmer jump (see Sect.~\ref{sec:balmer}), X-ray weakness from Compton-thick gas and non-Gaussian exponential wings of electron-scattering (see Sect.~\ref{sec:es}) and blue UV-optical spectra with higher-ionisation lines and without strong absorption signatures (see Sect.~\ref{sec:pop}). We address each of these predictions in turn. 

\subsection{Balmer-jump emission from nebular recombination}\label{sec:balmer}
Fig.~\ref{fig:Balmer_jump_com} shows that both \sirocco model spectra and selected LBD observations display a continuum offset across the Balmer limit. 
We quantify this feature by fitting a linear continuum, \(F_{\nu,\mathrm{lin}}(\lambda)\), to $F_\nu$ over the rest-frame interval \(2500<\lambda<3640\) Å, blueward of the Balmer edge, and extrapolating this fit to longer wavelengths. We define the continuum-normalised Balmer-jump offset as $\left\langle F_\nu/F_{\nu,\mathrm{lin}}\right\rangle$, evaluated in the relatively line-free redward windows \(4180\!-\!4280\) Å (redward of H$\delta$) and \(4510\!-\!4630\) Å (redward of H$\gamma$). Values \(\langle F_\nu/F_{\nu,\mathrm{lin}}\rangle<1\) indicate a redward continuum deficit, with smaller values corresponding to a stronger Balmer-jump signature.  
The spectra show a deficit redward of the edge, consistent with Balmer-jump emission from nebular recombination. The feature is not expected to be abrupt (e.g., Fig.~\ref{fig:Balmer_jump_com}, right panel), because broad emission from high-order hydrogen lines smooths the continuum, similar to Paschen jumps in LRDs \citep{Sneppen2026b}. \ion{Fe}{ii} pseudo-continuum emission could in principle mimic a similar continuum offset \citep{Ji2025b}; however, we find no direct evidence for the other \ion{Fe}{ii} line emission and the blue continuum often extends smoothly down to 2000\,\AA.

The LBDs shown in Fig.~\ref{fig:Balmer_jump_com} generally exhibit a Balmer-limit continuum offset in both line-free wavelength intervals. GN\,11836 and Nexus\,5819 show the most prominent and the most significant offset (see Table~\ref{tab:balmer_jump_offsets}). The uncertainty in each offset combines the statistical uncertainty on the mean flux within the wavelength window and the uncertainty propagated from the joint posterior of the continuum slope and intercept. At $2\sigma$ significance 10 out of 16 probed wavelength windows are in favour of a Balmer jump offset, while no object in any spectral window has significant evidence for a Balmer break (e.g., Balmer absorption). A shorter wavelength baseline provides less leverage on the fitted continuum slope, making the inferred significance particularly sensitive to the adopted fitting interval for the noisier spectra. To assess sensitivity to the linear continuum estimate, we repeated the fits using wavelength-intervals $\lambda\in[2200\,\AA,3640\,\AA]$ and $\lambda\in[2800\,\AA,3640\,\AA]$. Relative to the fiducial measurements, the mean absolute changes in \(\langle F_\nu/F_{\nu,\mathrm{lin}}\rangle\) are 0.03 and 0.09, respectively, while the number of individual measurements significant at \(2\sigma\) becomes eight and six, respectively. Balmer jump emission is thus likely common in the LBD population, but given the limited S/N and significance we cannot conclude (or indeed rule out) that it is ubiquitous. Regardless, in the model-space it is not expected that every object displays a Balmer jump given that the cocooned AGN population should bridge from Balmer emission LBD-like spectra into the Balmer absorption LRD-like spectra. 

\begin{table}
\caption{Continuum-normalised Balmer-jump offset measurements.}
\label{tab:balmer_jump_offsets}
\centering
\small
\begin{tabular*}{\columnwidth}
{@{\extracolsep{\fill}}lcccc@{}}
\hline\hline
&
\multicolumn{2}{c}{4180--4280\,\AA}
&
\multicolumn{2}{c}{4510--4630\,\AA}
\\
\cline{2-3}\cline{4-5}
Object
& Offset & Sig.
& Offset & Sig. \\
&
& ($\sigma$)
&
& ($\sigma$) \\
\hline
Nexus 5819
& $0.75\pm0.04$ & 6.1
& $0.66\pm0.04$ & 9.1 \\

Rubies 50052
& $0.84\pm0.14$ & 1.1
& $0.78\pm0.07$ & 3.1 \\

GS 3073
& $0.90\pm0.03$ & 4.1
& $0.96\pm0.03$ & 1.4 \\

GN 11836
& $0.51\pm0.04$ & 13.1
& $0.46\pm0.04$ & 14.8 \\

GN 2916
& $0.88\pm0.07$ & 1.6
& $1.09\pm0.13$ & $0.7$ \\

GS 179198
& $0.83\pm0.05$ & 3.0
& $0.81\pm0.07$ & 2.7 \\

GS 30148179
& $0.95\pm0.07$ & 0.8
& $0.88\pm0.05$ & 2.2 \\

GS 159438
& $0.93\pm0.04$ & 2.0
& $0.97\pm0.05$ & 0.7 \\
\hline
\end{tabular*}

\tablefoot{
The offset significance is
$|1-\langle F_\nu/F_{\nu,\mathrm{lin}}\rangle|/
\sigma_{\langle F_\nu/F_{\nu,\mathrm{lin}}\rangle}$;
all objects and windows with the exception of GN 2916 (at 4510--4630\,\AA) indicate flux below the extrapolated blue continuum. For each wavelength window, the quoted uncertainty combines in quadrature the pixel-to-pixel scatter of the continuum-normalised flux divided by $\sqrt{N}$ and the propagated uncertainty in the linear-continuum best-fit coefficients. 
}
\end{table}

\subsection{A shared electron-scattering regime: Compton-thick X-ray suppression and exponential line wings}\label{sec:es}
The presence of strong bound-free emission in LBDs and LRDs implies that a substantial ionised gas column exists around the central source. The same gas column both produces broad exponential-shaped line wings and suppresses X-ray emission, see Fig.~\ref{fig:Balmer_jump_pan}. The fact that all available fitted broad H$\alpha$ lines in LBDs are better or equally well modelled with exponential profiles rather than by multiple Gaussians \citep[e.g.][]{Rusakov2025}, in contrast to the general \jwst/X-ray AGN population \citep[e.g.][]{Scholtz2026}, provides strong evidence for an electron-scattering interpretation. This is further substantiated by the X-ray weakness of these objects, since such Compton-thick columns lead to significant X-ray suppression \citep[e.g.\ see the modest dependence on break strength for low column density models in Fig.~A1 of][]{Sneppen2026c}. Here, the left panel in Fig.~\ref{fig:Balmer_jump_pan} provides a consistency check: the column density regime that produces broad exponential H$\alpha$ wings also suppresses X-ray emission below current limits at $0.1\,Z_\odot$. The dearth of X-ray emission among LBDs may therefore constrain both their intrinsic X-ray emission and their obscuring gas columns in the same way as the X-ray weakness of LRDs: both require substantial attenuation by Compton-thick gas for an intrinsically AGN-like X-ray spectrum \citep{Maiolino_Chandra_2024,Comastri2026,Tortosa2026}.

\begin{figure}
\vspace{-0.1cm}
\begin{center}
    \includegraphics[angle=0,width=0.495\textwidth, viewport=2 22 405 520 ,clip=]{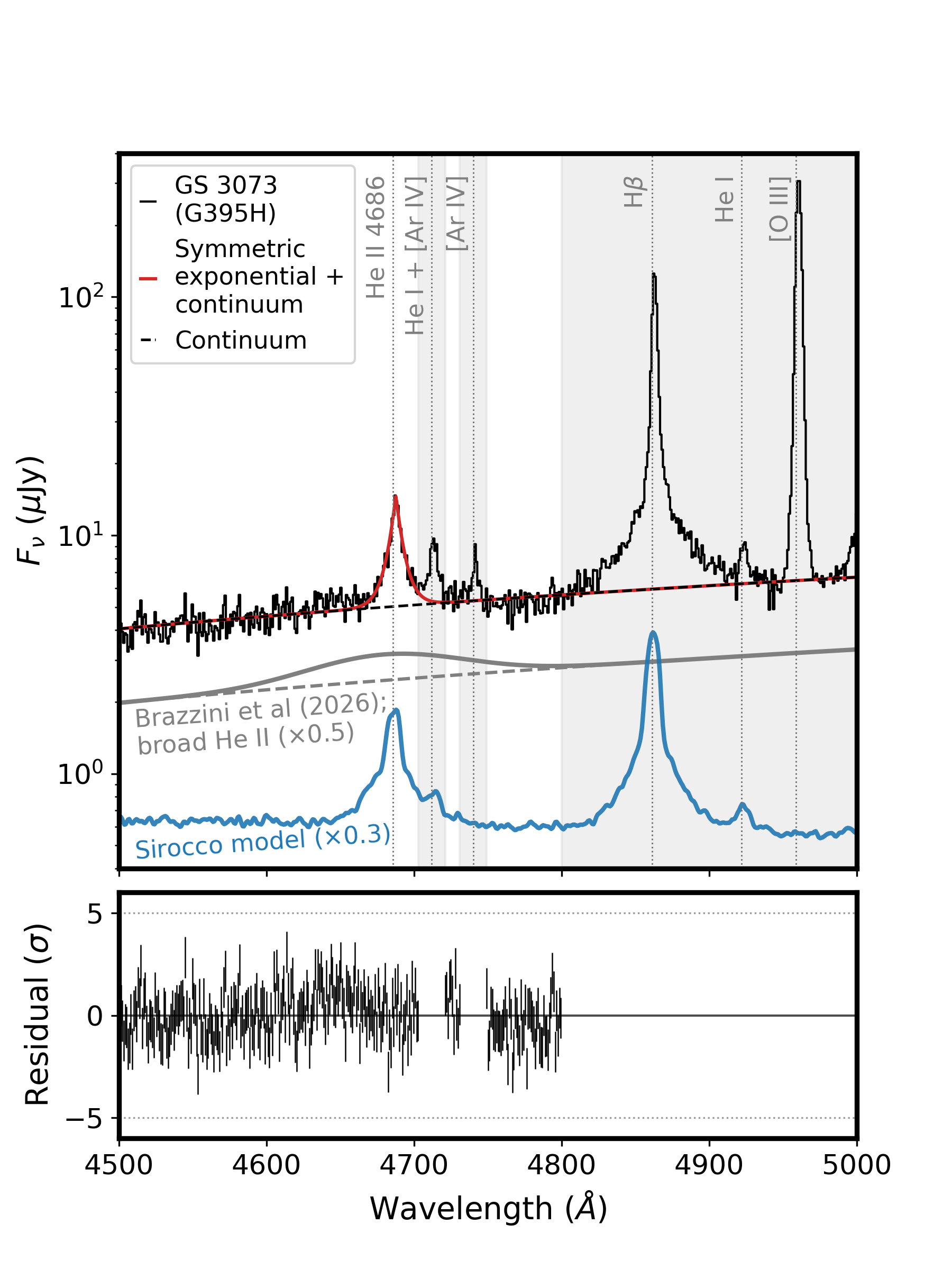}
\end{center}
\vspace{-0.5cm}
\caption{Comparison of the GS\,3073 spectrum and the \sirocco model around H$\beta$ and He\,II\,$\lambda4686$. The model reproduces the overall line profiles and relative strengths of the H, \hei, and \heii emission. For comparison, we show a best-fit exponential fit to He\,II\,$\lambda4686$ with residuals and the very broad \heii component inferred by \citet{Brazzini2026} from their three-component fit. In our single-column scattering geometry, the line profile instead naturally comprises two characteristic velocity scales (e.g., an unscattered core and electron-scattered wings). }\label{fig:HeII}
\end{figure}

\subsection{Predicted properties of cocoon models displaying Balmer jumps}\label{sec:pop}
\begin{figure}
\vspace{-0.3cm}
\begin{center}
    \includegraphics[angle=0,width=0.49\textwidth]{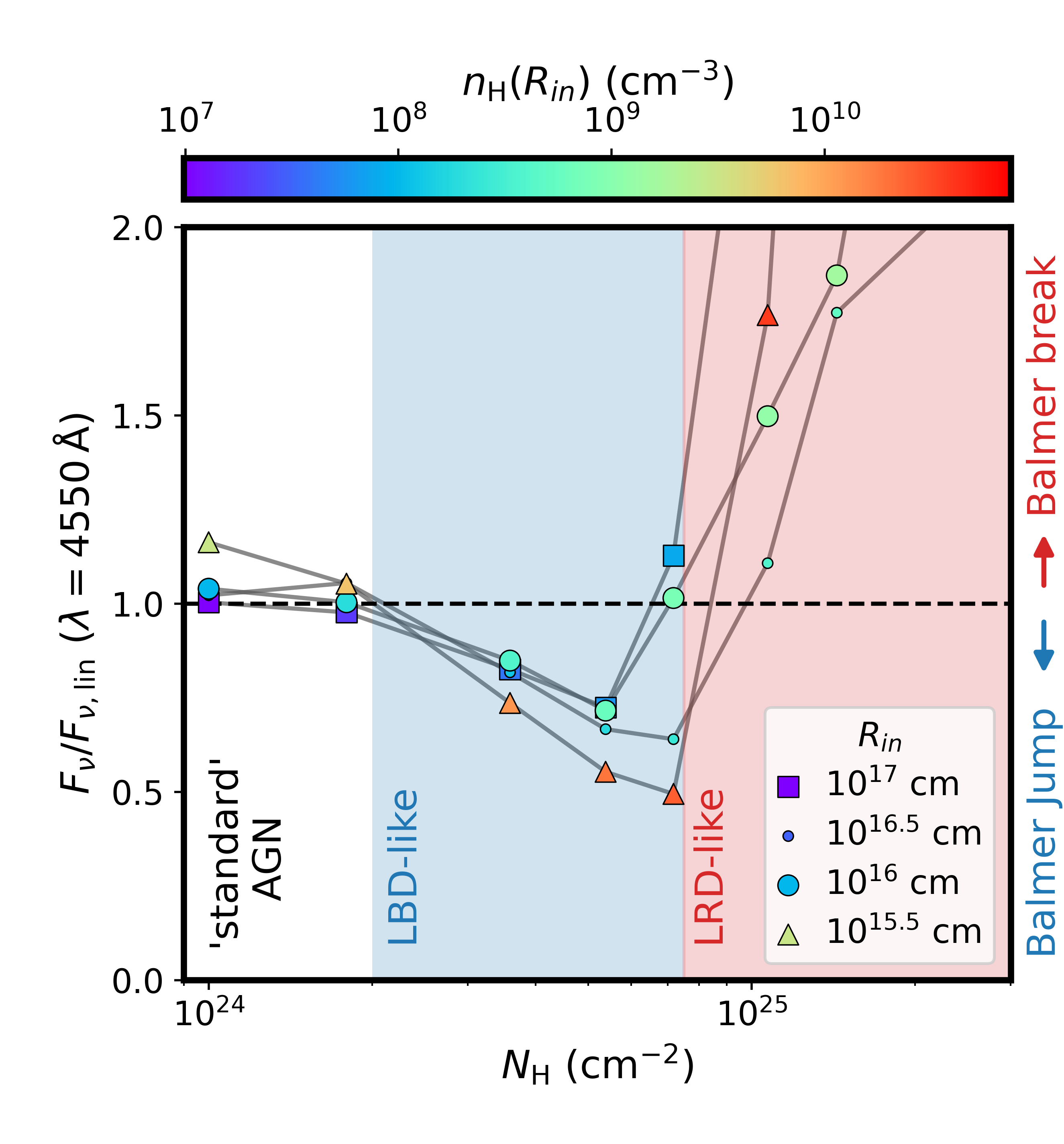}
\end{center}
\vspace{-0.9cm}
\caption{Balmer break/jump strength versus $N_H$ for various \sirocco density-sequences. Colour denotes the local hydrogen density at the inner boundary,
\(n_{\rm H}(R_{\rm in})\), while marker shape denotes
\(R_{\rm in}\) (large circles show the sequence plotted in Fig.~\ref{fig:dens_seq}). Values below unity indicate a Balmer jump, while values above unity indicate a Balmer break. Intermediate columns, $N_H\in(2-7)\times10^{24} {\rm cm^{-2}}$, produce LBD-like Balmer jumps, whereas higher columns can produce LRD-like Balmer breaks.  }
\label{fig:NH}
\end{figure}

The lower column density cocoon models with \sirocco allow several observational predictions to be made for objects with Balmer jumps. First, like LRDs, they should have broad lines with exponential wings due to electron scattering. Second, they should have larger core-to-broad line component flux-ratios and smaller broad-component FWHMs compared to LRDs (see Fig.~\ref{fig:dens_seq}, inset panel). Third, strong H and He absorption should be weaker or absent in LBDs. These trends arise because, along the increasing-column sequence, a modest but appreciable electron-scattering optical depth is reached before the partially neutral absorbing column becomes strong enough to imprint the characteristic LRD-like absorption spectrum. 
We note that the precise onset column density of absorption signatures depends on the ionisation structure of the cocoon. However, Fig.~\ref{fig:NH} shows that the emergence of the Balmer jump and the transition to Balmer breaks remain primarily organised by $N_{\rm H}$, across model sequences spanning both density normalisation and geometrical scale. Across the model space we have not found model spectra with simultaneous Balmer jumps and Balmer-line absorption, suggesting such observational configurations should be rare. 

\begin{figure}
\vspace{-0.1cm}
\begin{center}
    \includegraphics[angle=0,width=0.49\textwidth]{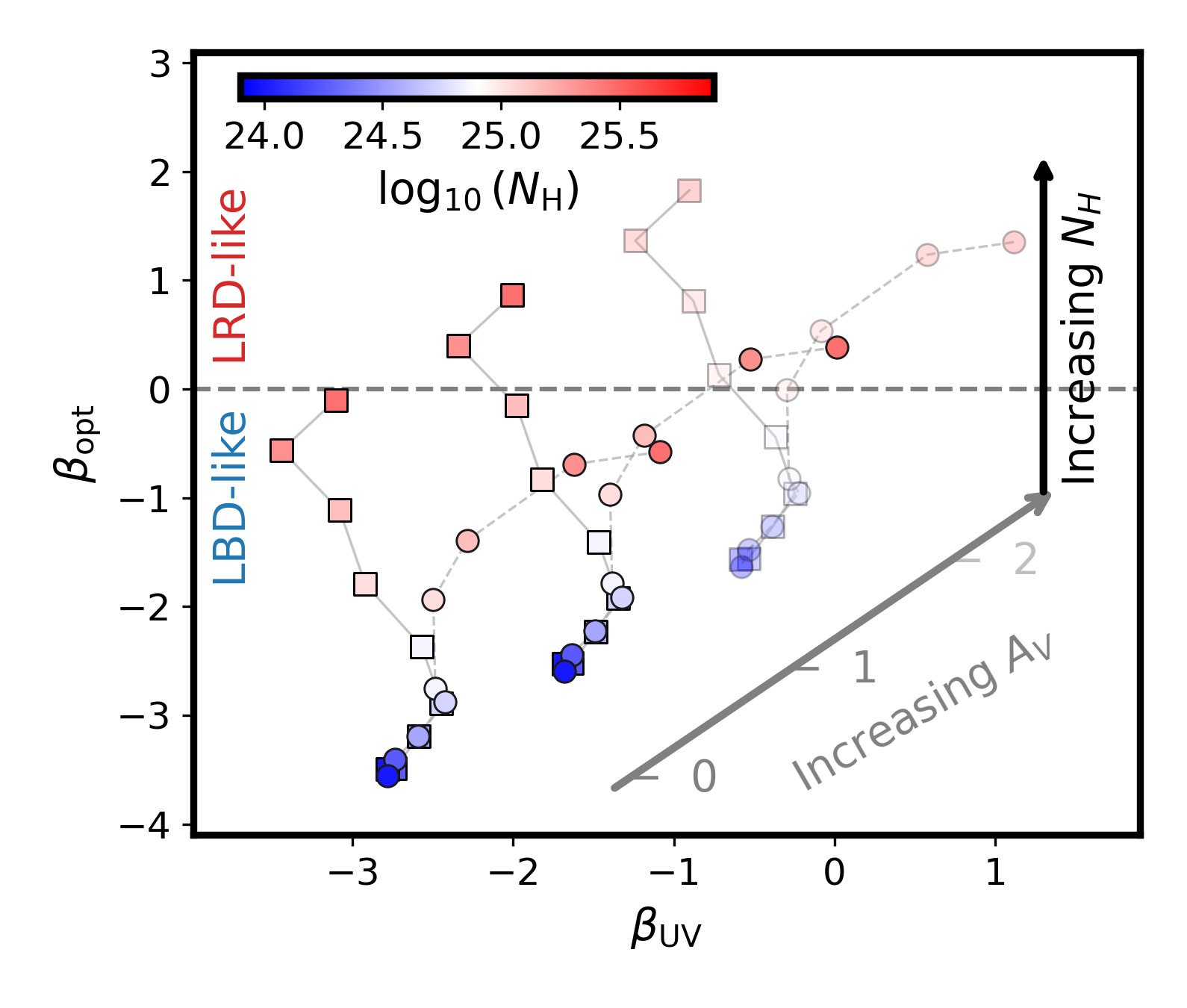}
\end{center}
\vspace{-0.9cm}
\caption{\sirocco model spectra optical ($\beta_{\rm opt}$) versus UV ($\beta_{\rm UV}$) continuum spectral slopes. 
The slopes $\beta_{\rm UV}$ and $\beta_{\rm opt}$ are measured using the rest-frame continuum windows in \cite{Brazzini2026} (i.e., 2160–3100 Å for $\beta_{\rm UV}$ and 5360–6270 Å for $\beta_{\rm opt}$, masking \ion{Mg}{ii}, \ion{He}{i}\,$\lambda$5876). Circles and squares show two cocoon density-sequences starting at $10^{16} {\rm cm}$ and $10^{17} {\rm cm}$, respectively. The dividing line highlights regions associated with LBDs and LRDs, where the predicted Balmer jump \sirocco models from Fig.~\ref{fig:NH} naturally inhabit this LBD color-space. }
\label{fig:beta}
\end{figure}

As a consequence of the limited absorption, these predicted Balmer-jump objects should display intrinsically blue spectra, with blue slopes in both the UV and optical. This naturally matches the colour-space suggested for LBDs (see Fig.~\ref{fig:beta}). While dust attenuation can redden both the UV and optical continua, increasing gas column density predominantly reddens the optical spectrum.

The \sirocco models allow additional LBD predictions to be made, which are more parameter-dependent than those indicated above. Because lower column density cocoons transmit more of the intrinsic optical continuum, H$\alpha$ is less tightly coupled to the underlying continuum and its luminosity and equivalent width can be reduced relative to more nebular-dominated LRD-like models. The largest equivalent widths, reaching ${\rm EW}\sim2000\,\AA$, occur near the transition between the LBD- and LRD-like regimes ($N_{\rm H}\sim10^{25}\,{\rm cm}^{-2}$; Fig.~\ref{fig:dens_seq}). Here the continuum is recombination dominated, while the column remains too low to substantially reprocess the line emission into continuum radiation.
If the parts of the gas experiencing a harder ionising field are no longer obscured, higher-ionisation lines in the UV and optical could become visible. For instance, He\,II\,$\lambda$4686/H$\beta$ increases as we move from LRD columns to LBD columns in Fig.~\ref{fig:dens_seq}. Indeed, such high-ionisation emission lines have been reported in GS\,3073 \citep[e.g.][]{Ubler2023,Ji2024}, while their detections and upper limits are informative in broader samples \citep[without separating LRDs from LBDs,][]{Lambrides2024} and in LRDs specifically \citep{Ando2026}. In an electron-scattering scenario, \heii emitted deeper within the scattering medium should undergo more scatterings and likely develop a broader profile. However, the claimed coexistence in GS\,3073 of an extremely broad component and narrow unscattered \heii emission \citep{Brazzini2026} is difficult to reproduce with a single isotropic scattering column (see Fig.~\ref{fig:HeII}). If such extremely broad features are robust, they may disfavour an electron-scattering scenario or instead point to multidimensional geometry containing a range of scattering columns. C\,III] and C\,IV emission are also often broadened (Figs.~\ref{fig:dens_seq}--\ref{fig:Balmer_jump_pan}), with the latter often exhibiting self-absorption. The model spectra also predict broad Ly$\alpha$ emission from the cocoon itself (e.g.\ Figs.~\ref{fig:dens_seq}-\ref{fig:Balmer_jump_pan}), a feature also expected in the higher column density \sirocco cocoon models \citep{Sneppen2026} and in observations of LRDs \citep{Ji2026,Ji2026b,Tang2026,Geris2026}. \cite{Geris2026} did find enhanced Ly-$\alpha$ equivalent widths in some individual LBDs, such as GS 3073, but did not establish the presence of broad Ly-$\alpha$ in individual LBDs or in their LBD stack. Regardless, given the Sobolev approximation assumed in \sirocco and without a dedicated Ly-$\alpha$ radiative-transfer code, we do not here attempt to quantify the emergent line shape and intensity.

\section{Discussion}
In \cite{Sneppen2026}, we showed that a range of column densities in AGN cocoons predicts a sequence in Balmer continuum emission/absorption, while the various outflows and inflows observed in individual objects could be explained with viewing-angle effects. A different orientation-based interpretation has recently been proposed by \cite{Madau2026}, in which LRDs are the dust-reddened, high-inclination counterparts of intrinsically blue compact broad-line AGN, or LBDs, powered by super-Eddington accretion. In their model, a geometrically thick accretion flow produces an anisotropic, blue ionising continuum, with self-shadowing suppressing the blue continuum along equatorial sightlines. Both interpretations connect LBDs and LRDs within one rapidly accreting AGN population. The distinction is that \cite{Madau2026} attribute the colour sequence primarily to orientation and self-shadowing in a thick accretion flow, whereas here the controlling parameter is the gas column through an electron-scattering cocoon \citep{Sneppen2026}. Similarly, \cite{Matthee2026} inferred a closely related picture, arguing that the transition from blue objects with centrally concentrated, absorption-free Balmer profiles to red objects with P-Cygni or absorption-dominated profiles correlates with Balmer-break strength, and interpreting this trend as primarily a sequence in column density. 

In our model, lower columns show recombination continua and LBD-like Balmer jumps, whereas higher columns introduce LRD-like Balmer-continuum and line absorption (see Fig.~\ref{fig:NH}). In this picture, LRDs and LBDs are not separate physical classes, but may instead trace different column densities, perhaps due to evolutionary stage, of dense gas surrounding rapidly growing black holes. The present calculations do not determine whether these column-density differences follow from temporal evolution of individual systems or object-to-object variation at comparable evolutionary stages.


\cite{Brazzini2026} identify GS\,3073 as a prototypical LBD and highlight properties shared by LBDs and LRDs, including compact morphologies, X-ray weakness and non-Gaussian broad Balmer-line profiles, which are all properties predicted for lower-column analogues of LRDs as shown here. GS\,3073 differs from LRDs in the lack of strong Balmer absorption and in having prominent \ion{He}{ii}\,$\lambda$4686 emission. This \ion{He}{ii} emission has a line component broader than H$\alpha$. In the picture developed here and explored with \sirocco , these differences are naturally associated with lower-column densities through the cocoon (see Fig.~\ref{fig:Balmer_jump_com}): the Balmer recombination continuum remains visible, strong neutral absorption is avoided, and \ion{He}{ii} emission for the more translucent gas can perhaps emerge. These higher-ionisation lines are broader than the hydrogen lines because they arise preferentially in the innermost parts of the cocoon \citep[e.g.][]{Chang2025}. Thus GS\,3073 provides a useful test because it exhibits both the LBD phenomenology emphasised by \cite{Brazzini2026} and the Balmer-jump behaviour predicted for the low-column extension of cocooned AGN models.

In this paper we have made a qualitative comparison to LBDs in general using a selected sample of eight LBD spectra to illustrate the point. This is because full Monte Carlo radiative-transfer calculations are computationally expensive, with a \sirocco spectrum requiring of order $10^2$ CPU hours, so we have focused here on presenting the general features and predictions of lower column density cocoon models. The natural next step -- extracting physical parameters by detailed model inference with LBD spectra, i.e. fitting individual LBD spectra, as well as comparing the population predictions against upcoming and future LBD samples -- will be addressed in future work.

\section{Conclusions}
\sirocco radiative transfer modelling, previously applied to LRDs, predicts a population of lower-column cocooned AGN, which are characterised by i) blue UV-optical colours, ii) compactness, iii) X-ray weakness, iv) exponential broad lines and v) Balmer-jump emission. Relative to LRDs, their broad permitted hydrogen lines should have narrower FWHM values, smaller luminosities, while their equivalent width distribution should retain a similarly high median and large variance, reaching EW$\sim$2000\,Å. A larger fraction of the line emission should also escape in the unscattered `core' component. Unlike their LRD counterparts, they should not be characterised by strong absorption signatures and they can include higher-ionisation lines such as \ion{He}{ii}\,$\lambda$4686. We suggest that the emerging population of little blue dots may correspond to this lower-column counterpart of LRDs. The LBD spectra highlighted here already show several of these predicted properties, and larger samples will provide a direct test of this interpretation.




\begin{acknowledgements}
The authors would like to thank Kenta Hotokezaka and Stuart Sim for helpful discussions. The \sirocco code \citep{Matthews2025} and documentation are publicly available\footnote{\href{https://github.com/sirocco-rt/sirocco}{https://github.com/sirocco-rt/sirocco}}. The comparison to observed LBDs makes use of the public \jwst data collected as part of several observational programs with the NIRSpec spectrograph \citep{Jakobsen2022} with PIDs: 1181, 1216, 1286, 4233, 5105. These observations have been uniformly reduced and published as part of the Dawn \jwst Archive\footnote{\href{https://dawn-cph.github.io/dja}{https://dawn-cph.github.io/dja}} (DJA), \cite{DeGraaff2024_RUBIES,Heintz2025,Pollock2025}. DJA is an initiative of the Cosmic Dawn Center (DAWN), which is funded by the Danish National Research Foundation under grant DNRF140. AS, DW \& GPN are funded in part by the European Union (ERC, HEAVYMETAL, 101071865). Views and opinions expressed are, however, those of the authors only and do not necessarily reflect those of the European Union or the European Research Council. Neither the European Union nor the granting authority can be held responsible for them. AS acknowledges support from Independent Research Fund Denmark (DFF) under grant 6219-00033B. JHM acknowledges funding from a Royal Society University Research Fellowship (URFR1221062).

\end{acknowledgements}



\bibliographystyle{aa}
\bibliography{refs} 

\begin{thebibliography}{45}
\expandafter\ifx\csname natexlab\endcsname\relax\def\natexlab#1{#1}\fi

\bibitem[{{Ando} {et~al.}(2026){Ando}, {Harikane}, {Katz}, {Inayoshi}, \& {Tanaka}}]{Ando2026}
{Ando}, M., {Harikane}, Y., {Katz}, H., {Inayoshi}, K., \& {Tanaka}, T.~S. 2026, arXiv e-prints, arXiv:2606.03522

\bibitem[{{Brazzini} {et~al.}(2026){Brazzini}, {D'Eugenio}, {Maiolino}, {Lyu}, {DeCoursey}, {{\"U}bler}, {Ji}, {Juod{\v{z}}balis}, {Scholtz}, {Jones}, {Hainline}, {Dalla Bont{\`a}}, {{\'e}rez-Gonz{\'a}lez}, {Geris}, {Harshan}, {Feruglio}, {Bischetti}, {Mazzolari}, {Rieke}, {Alberts}, {Trefoloni}, {Carniani}, {Parlanti}, {Marconi}, {Risaliti}, {Ramos Almeida}, {Rinaldi}, {Perna}, {Zamora}, {Lamperti}, {Venturi}, {Cresci}, {Bunker}, \& {Ivey}}]{Brazzini2026}
{Brazzini}, M., {D'Eugenio}, F., {Maiolino}, R., {et~al.} 2026, arXiv e-prints, arXiv:2601.22214

\bibitem[{{Brooks} {et~al.}(2024){Brooks}, {Simons}, {Trump}, {Taylor}, {Backhaus}, {Davis}, {Buat}, {Cleri}, {Finkelstein}, {Hirschmann}, {Holwerda}, {Kocevski}, {Koekemoer}, {Lucas}, {Pacucci}, \& {Seill{\'e}}}]{Brooks2024}
{Brooks}, M., {Simons}, R.~C., {Trump}, J.~R., {et~al.} 2024, arXiv e-prints, arXiv:2410.07340

\bibitem[{{Calzetti} {et~al.}(2000){Calzetti}, {Armus}, {Bohlin}, {Kinney}, {Koornneef}, \& {Storchi-Bergmann}}]{Calzetti2000}
{Calzetti}, D., {Armus}, L., {Bohlin}, R.~C., {et~al.} 2000, \apj, 533, 682

\bibitem[{{Chang} {et~al.}(2026){Chang}, {Gronke}, {Matthee}, \& {Mason}}]{Chang2025}
{Chang}, S.-J., {Gronke}, M., {Matthee}, J., \& {Mason}, C. 2026, \mnras, 545, staf2131

\bibitem[{{Chen} {et~al.}(2025){Chen}, {Li}, {Inayoshi}, \& {Ho}}]{Chen2025}
{Chen}, K., {Li}, Z., {Inayoshi}, K., \& {Ho}, L.~C. 2025, \apjl, 994, L42

\bibitem[{{Comastri} {et~al.}(2026){Comastri}, {Lanzuisi}, {Vito}, {Marchesi}, {Brusa}, {Gilli}, {Juod{\v{z}}balis}, {Maiolino}, {Mazzolari}, {Risaliti}, {Scholtz}, \& {Vignali}}]{Comastri2026}
{Comastri}, A., {Lanzuisi}, G., {Vito}, F., {et~al.} 2026, \aap, 706, A302

\bibitem[{{de Graaff} {et~al.}(2025{\natexlab{a}}){de Graaff}, {Brammer}, {Weibel}, {Lewis}, {Maseda}, {Oesch}, {Bezanson}, {Boogaard}, {Cleri}, {Cooper}, {Gottumukkala}, {Greene}, {Hirschmann}, {Hviding}, {Katz}, {Labb{\'e}}, {Leja}, {Matthee}, {McConachie}, {Miller}, {Naidu}, {Price}, {Rix}, {Setton}, {Suess}, {Wang}, {Whitaker}, \& {Williams}}]{DeGraaff2024_RUBIES}
{de Graaff}, A., {Brammer}, G., {Weibel}, A., {et~al.} 2025{\natexlab{a}}, \aap, 697, A189

\bibitem[{{de Graaff} {et~al.}(2025{\natexlab{b}}){de Graaff}, {Rix}, {Naidu}, {Labb{\'e}}, {Wang}, {Leja}, {Matthee}, {Katz}, {Greene}, {Hviding}, {Baggen}, {Bezanson}, {Boogaard}, {Brammer}, {Dayal}, {van Dokkum}, {Goulding}, {Hirschmann}, {Maseda}, {McConachie}, {Miller}, {Nelson}, {Oesch}, {Setton}, {Shivaei}, {Weibel}, {Whitaker}, \& {Williams}}]{deGraaff2025}
{de Graaff}, A., {Rix}, H.-W., {Naidu}, R.~P., {et~al.} 2025{\natexlab{b}}, \aap, 701, A168

\bibitem[{{D'Eugenio} {et~al.}(2025){D'Eugenio}, {Nelson}, {Ji}, {Baggen}, {Greene}, {Labb{\'e}}, {Pezzulli}, {Brown}, {Maiolino}, {Matthee}, {Terlevich}, {Terlevich}, {Torralba}, \& {Carniani}}]{DEugenio2025b}
{D'Eugenio}, F., {Nelson}, E., {Ji}, X., {et~al.} 2025, arXiv e-prints, arXiv:2510.00101

\bibitem[{{Geris} {et~al.}(2026){Geris}, {Maiolino}, {Ji}, {Risaliti}, {Lanzuisi}, {D'Eugenio}, {Isobe}, {Jones}, {Harshan}, {Brazzini}, {Juod{\v{z}}balis}, {Scholtz}, {Rinaldi}, {{\"U}bler}, {Baker}, {Bunker}, {Brusa}, {Carniani}, {Charlot}, {Curti}, {Comastri}, {Lake}, {Gilli}, {Hainline}, {Madau}, {Marchesi}, {Mazzolari}, {Napolitano}, {Parlanti}, {Pentericci}, {Ramos Almeida}, {Robertson}, {Silcock}, {Tripodi}, {Venturi}, {Vignali}, {Vito}, \& {Zhu}}]{Geris2026}
{Geris}, S., {Maiolino}, R., {Ji}, X., {et~al.} 2026, arXiv e-prints, arXiv:2606.21614

\bibitem[{{Harikane} {et~al.}(2023){Harikane}, {Zhang}, {Nakajima}, {Ouchi}, {Isobe}, {Ono}, {Hatano}, {Xu}, \& {Umeda}}]{Harikane2023}
{Harikane}, Y., {Zhang}, Y., {Nakajima}, K., {et~al.} 2023, \apj, 959, 39

\bibitem[{{Heintz} {et~al.}(2025){Heintz}, {Brammer}, {Watson}, {Oesch}, {Keating}, {Hayes}, {Abdurro'uf}, {Arellano-C{\'o}rdova}, {Carnall}, {Christiansen}, {Cullen}, {Dav{\'e}}, {Dayal}, {Ferrara}, {Finlator}, {Fynbo}, {Flury}, {Gelli}, {Gillman}, {Gottumukkala}, {Gould}, {Greve}, {Hardin}, {Hsiao}, {Hutter}, {Jakobsson}, {Killi}, {Khosravaninezhad}, {Laursen}, {Lee}, {Magdis}, {Matthee}, {Naidu}, {Narayanan}, {Pollock}, {Prescott}, {Rusakov}, {Shuntov}, {Sneppen}, {Smit}, {Tanvir}, {Terp}, {Toft}, {Valentino}, {Vijayan}, {Weaver}, {Wise}, \& {Witstok}}]{Heintz2025}
{Heintz}, K.~E., {Brammer}, G.~B., {Watson}, D., {et~al.} 2025, \aap, 693, A60

\bibitem[{{Inayoshi} \& {Maiolino}(2025)}]{Inayoshi2025}
{Inayoshi}, K. \& {Maiolino}, R. 2025, \apjl, 980, L27

\bibitem[{{Jakobsen} {et~al.}(2022){Jakobsen}, {Ferruit}, {Alves de Oliveira}, {Arribas}, {Bagnasco}, {Barho}, {Beck}, {Birkmann}, {B{\"o}ker}, {Bunker}, {Charlot}, {de Jong}, {de Marchi}, {Ehrenwinkler}, {Falcolini}, {Fels}, {Franx}, {Franz}, {Funke}, {Giardino}, {Gnata}, {Holota}, {Honnen}, {Jensen}, {Jentsch}, {Johnson}, {Jollet}, {Karl}, {Kling}, {K{\"o}hler}, {Kolm}, {Kumari}, {Lander}, {Lemke}, {L{\'o}pez-Caniego}, {L{\"u}tzgendorf}, {Maiolino}, {Manjavacas}, {Marston}, {Maschmann}, {Maurer}, {Messerschmidt}, {Moseley}, {Mosner}, {Mott}, {Muzerolle}, {Pirzkal}, {Pittet}, {Plitzke}, {Posselt}, {Rapp}, {Rauscher}, {Rawle}, {Rix}, {R{\"o}del}, {Rumler}, {Sabbi}, {Salvignol}, {Schmid}, {Sirianni}, {Smith}, {Strada}, {te Plate}, {Valenti}, {Wettemann}, {Wiehe}, {Wiesmayer}, {Willott}, {Wright}, {Zeidler}, \& {Zincke}}]{Jakobsen2022}
{Jakobsen}, P., {Ferruit}, P., {Alves de Oliveira}, C., {et~al.} 2022, \aap, 661, A80

\bibitem[{{Ji} {et~al.}(2025){Ji}, {Maiolino}, {Ferland}, {D'Eugenio}, {Bhatawdekar}, {Charlot}, {Chevallard}, {Curti}, {Curtis-Lake}, {Hainline}, {Ji}, {Robertson}, {Rodr{\'\i}guez Del Pino}, {Scholtz}, {Tacchella}, {Williams}, \& {Witstok}}]{Ji2025b}
{Ji}, X., {Maiolino}, R., {Ferland}, G., {et~al.} 2025, \mnras, 541, 2134

\bibitem[{{Ji} {et~al.}(2026{\natexlab{a}}){Ji}, {Pezzulli}, {D'Eugenio}, {Maiolino}, {Carniani}, {Tacchella}, {Jones}, {Smith}, {Witstok}, {Fabian}, {Geris}, {Harshan}, {Isobe}, {Ivey}, {Juod{\v{z}}balis}, {Pascalau}, {Scholtz}, \& {Witten}}]{Ji2026}
{Ji}, X., {Pezzulli}, G., {D'Eugenio}, F., {et~al.} 2026{\natexlab{a}}, arXiv e-prints, arXiv:2604.03370

\bibitem[{{Ji} {et~al.}(2024){Ji}, {{\"U}bler}, {Maiolino}, {D'Eugenio}, {Arribas}, {Bunker}, {Charlot}, {Perna}, {Rodr{\'\i}guez Del Pino}, {B{\"o}ker}, {Cresci}, {Curti}, {Kumari}, \& {Lamperti}}]{Ji2024}
{Ji}, X., {{\"U}bler}, H., {Maiolino}, R., {et~al.} 2024, \mnras, 535, 881

\bibitem[{{Ji} {et~al.}(2026{\natexlab{b}}){Ji}, {Sun}, {Giavalisco}, {de Graaff}, {Williams}, {Zhu}, {Rieke}, \& {Rieke}}]{Ji2026b}
{Ji}, Z., {Sun}, Y., {Giavalisco}, M., {et~al.} 2026{\natexlab{b}}, arXiv e-prints, arXiv:2606.09970

\bibitem[{{Kocevski} {et~al.}(2023){Kocevski}, {Onoue}, {Inayoshi}, {Trump}, {Arrabal Haro}, {Grazian}, {Dickinson}, {Finkelstein}, {Kartaltepe}, {Hirschmann}, {Aird}, {Holwerda}, {Fujimoto}, {Juneau}, {Amor{\'\i}n}, {Backhaus}, {Bagley}, {Barro}, {Bell}, {Bisigello}, {Calabr{\`o}}, {Cleri}, {Cooper}, {Ding}, {Grogin}, {Ho}, {Hutchison}, {Inoue}, {Jiang}, {Jones}, {Koekemoer}, {Li}, {Li}, {McGrath}, {Molina}, {Papovich}, {P{\'e}rez-Gonz{\'a}lez}, {Pirzkal}, {Wilkins}, {Yang}, \& {Yung}}]{Kocevski2023_first}
{Kocevski}, D.~D., {Onoue}, M., {Inayoshi}, K., {et~al.} 2023, \apjl, 954, L4

\bibitem[{{Kokorev} {et~al.}(2025){Kokorev}, {Chisholm}, {Naidu}, {Fujimoto}, {Atek}, {Brammer}, {Finkelstein}, {Akins}, {Berg}, {Furtak}, {Fei}, {Hsiao}, {Matthee}, {Mu{\~n}oz}, {Oesch}, {Pan}, {Rinaldi}, {Saldana-Lopez}, {Schaerer}, {Volonteri}, \& {Zitrin}}]{Kokorev2025}
{Kokorev}, V., {Chisholm}, J., {Naidu}, R.~P., {et~al.} 2025, arXiv e-prints, arXiv:2511.07515

\bibitem[{{Lambrides} {et~al.}(2026){Lambrides}, {Larson}, {Garofali}, {Ptak}, {Chiaberge}, {Long}, {Hutchison}, {Norman}, {McKinney}, {Akins}, {Berg}, {Chisholm}, {Civano}, {Cloonan}, {Endsley}, {Faisst}, {Gilli}, {Gillman}, {Hirschmann}, {Kartaltepe}, {Kocevski}, {Kokorev}, {Pacucci}, {Richardson}, {Stiavelli}, \& {Whalen}}]{Lambrides2024}
{Lambrides}, E., {Larson}, R.~L., {Garofali}, K., {et~al.} 2026, Nature Astronomy [\eprint[arXiv]{2409.13047}]

\bibitem[{{Long} \& {Knigge}(2002)}]{long2002}
{Long}, K.~S. \& {Knigge}, C. 2002, \apj, 579, 725

\bibitem[{{Madau} \& {Maiolino}(2026)}]{Madau2026}
{Madau}, P. \& {Maiolino}, R. 2026, arXiv e-prints, arXiv:2605.05074

\bibitem[{{Maiolino} {et~al.}(2025){Maiolino}, {Risaliti}, {Signorini}, {Trefoloni}, {Juod{\v{z}}balis}, {Scholtz}, {{\"U}bler}, {D'Eugenio}, {Carniani}, {Fabian}, {Ji}, {Mazzolari}, {Bertola}, {Brusa}, {Bunker}, {Charlot}, {Comastri}, {Cresci}, {DeCoursey}, {Egami}, {Fiore}, {Gilli}, {Perna}, {Tacchella}, \& {Venturi}}]{Maiolino_Chandra_2024}
{Maiolino}, R., {Risaliti}, G., {Signorini}, M., {et~al.} 2025, \mnras [\eprint[arXiv]{2405.00504}]

\bibitem[{{Matthee} {et~al.}(2024){Matthee}, {Naidu}, {Brammer}, {Chisholm}, {Eilers}, {Goulding}, {Greene}, {Kashino}, {Labbe}, {Lilly}, {Mackenzie}, {Oesch}, {Weibel}, {Wuyts}, {Xiao}, {Bordoloi}, {Bouwens}, {van Dokkum}, {Illingworth}, {Kramarenko}, {Maseda}, {Mason}, {Meyer}, {Nelson}, {Reddy}, {Shivaei}, {Simcoe}, \& {Yue}}]{Matthee2024}
{Matthee}, J., {Naidu}, R.~P., {Brammer}, G., {et~al.} 2024, \apj, 963, 129

\bibitem[{{Matthee} {et~al.}(2026){Matthee}, {Torralba}, {Pezzulli}, {Naidu}, {Chisholm}, {Mascia}, {Greene}, {Ishikawa}, {Gronke}, {Wuyts}, {Bordoloi}, {Brammer}, {Chang}, {Eilers}, {de Graaff}, {Hviding}, {Iani}, {Illingworth}, {Kashino}, {Labbe}, {Ma}, {Maseda}, {Meyer}, {Nelson}, {Oesch}, \& {Xiao}}]{Matthee2026}
{Matthee}, J., {Torralba}, A., {Pezzulli}, G., {et~al.} 2026, arXiv e-prints, arXiv:2603.17667

\bibitem[{{Matthews} {et~al.}(2025){Matthews}, {Long}, {Knigge}, {Sim}, {Parkinson}, {Higginbottom}, {Mangham}, {Scepi}, {Wallis}, {Hewitt}, \& {Mosallanezhad}}]{Matthews2025}
{Matthews}, J.~H., {Long}, K.~S., {Knigge}, C., {et~al.} 2025, \mnras, 536, 879

\bibitem[{{Naidu} {et~al.}(2025){Naidu}, {Matthee}, {Katz}, {de Graaff}, {Oesch}, {Smith}, {Greene}, {Brammer}, {Weibel}, {Hviding}, {Chisholm}, {Labb{\'e}}, {Simcoe}, {Witten}, {Atek}, {Baggen}, {Belli}, {Bezanson}, {Boogaard}, {Bose}, {Covelo-Paz}, {Dayal}, {Fudamoto}, {Furtak}, {Giovinazzo}, {Goulding}, {Gronke}, {Heintz}, {Hirschmann}, {Illingworth}, {Inoue}, {Johnson}, {Leja}, {Leonova}, {McConachie}, {Maseda}, {Natarajan}, {Nelson}, {Setton}, {Shivaei}, {Sobral}, {Stefanon}, {Tacchella}, {Toft}, {Torralba}, {van Dokkum}, {van der Wel}, {Volonteri}, {Walter}, {Wang}, \& {Watson}}]{Naidu2025}
{Naidu}, R.~P., {Matthee}, J., {Katz}, H., {et~al.} 2025, arXiv e-prints, arXiv:2503.16596

\bibitem[{{Nikopoulos} {et~al.}(2025){Nikopoulos}, {Watson}, {Sneppen}, {Rusakov}, {Heintz}, {Witstok}, \& {Brammer}}]{Nikopoulos2025}
{Nikopoulos}, G.~P., {Watson}, D., {Sneppen}, A., {et~al.} 2025, arXiv e-prints, arXiv:2510.06362

\bibitem[{{Pollock} {et~al.}(2026){Pollock}, {Gottumukkala}, {Heintz}, {Brammer}, {Roberts-Borsani}, {Oesch}, {Witstok}, {Arellano-C{\'o}rdova}, {Cullen}, {Scholte}, {Terp}, {Rowland}, {Sneppen}, {Ito}, {Valentino}, {Matthee}, {Watson}, \& {Toft}}]{Pollock2025}
{Pollock}, C.~L., {Gottumukkala}, R., {Heintz}, K.~E., {et~al.} 2026, \aap, 708, A203

\bibitem[{{Rusakov} {et~al.}(2026){Rusakov}, {Watson}, {Nikopoulos}, {Brammer}, {Gottumukkala}, {Harvey}, {Heintz}, {Damgaard}, {Sim}, {Sneppen}, {Vijayan}, {Adams}, {Austin}, {Conselice}, {Goolsby}, {Toft}, \& {Witstok}}]{Rusakov2025}
{Rusakov}, V., {Watson}, D., {Nikopoulos}, G.~P., {et~al.} 2026, \nat, 649, 574

\bibitem[{{Scholtz} {et~al.}(2026){Scholtz}, {D'Eugenio}, {Maiolino}, {Brazzini}, {{\"U}bler}, {Ji}, {Perna}, {Sun}, {Brocchi}, {Carniani}, {Cresci}, {Ivey}, {Juod{\v{z}}balis}, {Marconi}, {Mazzolari}, {Risaliti}, \& {Trefoloni}}]{Scholtz2026}
{Scholtz}, J., {D'Eugenio}, F., {Maiolino}, R., {et~al.} 2026, arXiv e-prints, arXiv:2603.22277

\bibitem[{{Setton} {et~al.}(2025){Setton}, {Greene}, {de Graaff}, {Ma}, {Leja}, {Matthee}, {Bezanson}, {Boogaard}, {Cleri}, {Katz}, {Labbe}, {Maseda}, {McConachie}, {Miller}, {Price}, {Suess}, {van Dokkum}, {Wang}, {Weibel}, {Whitaker}, \& {Williams}}]{Setton2024}
{Setton}, D.~J., {Greene}, J.~E., {de Graaff}, A., {et~al.} 2025, \apj, 995, 118

\bibitem[{{Shen} {et~al.}(2024){Shen}, {Zhuang}, {Li}, {Burgasser}, {Fan}, {Greene}, {Narayan}, {Shapley}, {Sun}, {Wang}, \& {Yang}}]{Shen2024}
{Shen}, Y., {Zhuang}, M.-Y., {Li}, J., {et~al.} 2024, arXiv e-prints, arXiv:2408.12713

\bibitem[{{Sneppen} {et~al.}(2026{\natexlab{a}}){Sneppen}, {Matthews}, {Watson}, {Cameron}, {Sim}, {Witstok}, {Brammer}, {Heintz}, \& {Nikopoulos}}]{Sneppen2026b}
{Sneppen}, A., {Matthews}, J.~H., {Watson}, D., {et~al.} 2026{\natexlab{a}}, arXiv e-prints, arXiv:2604.09399

\bibitem[{{Sneppen} {et~al.}(2026{\natexlab{b}}){Sneppen}, {Watson}, {Matthews}, {Nikopoulos}, {Allen}, {Brammer}, {Damgaard}, {Heintz}, {Knigge}, {Long}, {Rusakov}, {Sim}, \& {Witstok}}]{Sneppen2026}
{Sneppen}, A., {Watson}, D., {Matthews}, J.~H., {et~al.} 2026{\natexlab{b}}, arXiv e-prints, arXiv:2601.18864

\bibitem[{{Sneppen} {et~al.}(2026{\natexlab{c}}){Sneppen}, {Watson}, {Matthews}, \& {Sim}}]{Sneppen2026c}
{Sneppen}, A., {Watson}, D., {Matthews}, J.~H., \& {Sim}, S.~A. 2026{\natexlab{c}}, arXiv e-prints, arXiv:2605.15263

\bibitem[{{Tang} {et~al.}(2026){Tang}, {Stark}, {Mason}, {Chen}, {Katz}, {Gronke}, {Furtak}, {Chang}, {Matthee}, {Whitler}, {Zitrin}, {Endsley}, {Gelli}, {Roychowdhury}, {Senchyna}, {Topping}, \& {Zhang}}]{Tang2026}
{Tang}, M., {Stark}, D.~P., {Mason}, C.~A., {et~al.} 2026, arXiv e-prints, arXiv:2604.03563

\bibitem[{{Torralba} {et~al.}(2025){Torralba}, {Matthee}, {Pezzulli}, {Naidu}, {Ishikawa}, {Brammer}, {Chang}, {Chisholm}, {de Graaff}, {D'Eugenio}, {Di Cesare}, {Eilers}, {Greene}, {Gronke}, {Iani}, {Kokorev}, {Kotiwale}, {Kramarenko}, {Ma}, {Mascia}, {Navarrete}, {Nelson}, {Oesch}, {Simcoe}, \& {Wuyts}}]{Torralba2025}
{Torralba}, A., {Matthee}, J., {Pezzulli}, G., {et~al.} 2025, arXiv e-prints, arXiv:2510.00103

\bibitem[{{Torralba} {et~al.}(2026){Torralba}, {Matthee}, {Weibel}, {Naidu}, {Ma}, {Cloonan}, {Desai}, {de Graaff}, {Greene}, {Jespersen}, {Kramarenko}, {Mascia}, {Oesch}, {Sun}, \& {Williams}}]{Torralba2026}
{Torralba}, A., {Matthee}, J., {Weibel}, A., {et~al.} 2026, arXiv e-prints, arXiv:2603.28335

\bibitem[{{Tortosa} {et~al.}(2026){Tortosa}, {Ricci}, {Du}, {Venturi}, {Ho}, {Li}, {Wang}, \& {Berton}}]{Tortosa2026}
{Tortosa}, A., {Ricci}, C., {Du}, P., {et~al.} 2026, arXiv e-prints, arXiv:2603.10162

\bibitem[{{{\"U}bler} {et~al.}(2023){{\"U}bler}, {Maiolino}, {Curtis-Lake}, {P{\'e}rez-Gonz{\'a}lez}, {Curti}, {Perna}, {Arribas}, {Charlot}, {Marshall}, {D'Eugenio}, {Scholtz}, {Bunker}, {Carniani}, {Ferruit}, {Jakobsen}, {Rix}, {Rodr{\'\i}guez Del Pino}, {Willott}, {Boeker}, {Cresci}, {Jones}, {Kumari}, \& {Rawle}}]{Ubler2023}
{{\"U}bler}, H., {Maiolino}, R., {Curtis-Lake}, E., {et~al.} 2023, \aap, 677, A145

\bibitem[{{Yanagisawa} {et~al.}(2026){Yanagisawa}, {Ouchi}, {Golubchik}, {Oguri}, {Fujimoto}, {Kokorev}, {Brammer}, {Sun}, {Nakane}, {Harikane}, {Umeda}, {Akins}, {Atek}, {Bauer}, {Brada{\v{c}}}, {Chisholm}, {Coe}, {Diego}, {Ferguson}, {Finkelstein}, {Furtak}, {Inayoshi}, {Koekemoer}, {Matthee}, {Naidu}, {Ono}, {Pan}, {Richard}, {Robbins}, {Willott}, {Zitrin}, {Amor{\'\i}n}, {Bradley}, {Bromm}, {Conselice}, {Dayal}, {Kartaltepe}, {Lopes}, {Lucas}, {Magdis}, {Martis}, {Papovich}, {Schaerer}, {Valentino}, {Vanzella}, {Allingham}, {Grogin}, {Gonz{\'a}lez-Otero}, {Ricotti}, \& {Windhorst}}]{Yanagisawa2026}
{Yanagisawa}, H., {Ouchi}, M., {Golubchik}, M., {et~al.} 2026, arXiv e-prints, arXiv:2601.06015

\bibitem[{{Zhuang} {et~al.}(2026){Zhuang}, {Shen}, {Li}, {Pan}, {Hu}, {Burgasser}, {Coulter}, {Greene}, \& {Wang}}]{Zhuang2026}
{Zhuang}, M.-Y., {Shen}, Y., {Li}, J., {et~al.} 2026, arXiv e-prints, arXiv:2603.04586

\end{thebibliography}

\setcounter{section}{0}
\setcounter{equation}{0}
\setcounter{figure}{0}
\renewcommand{\thesection}{Appendix \arabic{section}}
\renewcommand{\theequation}{A.\arabic{equation}}
\renewcommand{\thefigure}{A.\arabic{figure}}
\renewcommand{\thetable}{A.\arabic{table}}

\end{document}